\documentclass{article}

\usepackage{arxiv}

\usepackage[utf8]{inputenc} 
\usepackage[T1]{fontenc}    
\usepackage{hyperref}       
\usepackage{url}            
\usepackage{booktabs}       
\usepackage{amsfonts}       
\usepackage{nicefrac}       
\usepackage{microtype}      
\usepackage{lipsum}		
\usepackage{graphicx}
\usepackage{natbib}
\usepackage{doi}
\usepackage{amsmath}

\title{Statistical mechanics of beam self-cleaning in GRIN multimode optical fibers}

\author{F. Mangini$^{1,2,\dag,*}$, M. Gervaziev$^{3,\dag}$, M. Ferraro$^{2,\dag}$, D.S. Kharenko$^3$, M. Zitelli$^2$, Y. Sun$^2$, V. Couderc$^4$,\\ 
\textbf{E.V. Podivilov$^3$, S.A. Babin$^3$, and S. Wabnitz$^{2,3}$}\\
\\
    $^1$ Department of Information Engineering, University of Brescia, Brescia, Italy\\
	$^2$ Department of Information Engineering, Electronics, and Telecommunications, \\Sapienza University of Rome, Rome, Italy \\
	$^3$ Novosibirsk State University, Novosibirsk, Russia\\
	$^4$ Universit\'e de Limoges, XLIM, UMR CNRS, Limoges, France\\
	$^*$fabio.mangini@unibs.it\\
	$^\dag$These authors have contributed equally
	}
  
\begin{document}
\maketitle

\begin{abstract}
Since its first demonstration in graded-index multimode fibers, spatial beam self-cleaning has attracted a growing research interest. It allows for the propagation of beams with a bell-shaped spatial profile, thus enabling the use of multimode fibers 
for several applications, from biomedical imaging 
to high-power beam delivery. So far, beam self-cleaning has been experimentally studied under several different experimental conditions. Whereas it has been theoretically described as the irreversible energy transfer from high-order modes towards the fundamental mode, in analogy with a beam condensation mechanism. Here, we provide a definitive theoretical description of beam self-cleaning, by means of a semi-classical statistical mechanics model of wave thermalization. This approach is confirmed by an extensive experimental characterization, based on a holographic mode decomposition technique, employing laser pulses with temporal durations ranging from femtoseconds up to nanoseconds. An excellent agreement between theory and experiments is found, which demonstrates that beam self-cleaning can be fully described in terms of the basic conservation laws of statistical mechanics.
\end{abstract}


\section{Introduction}

Interest in optical multimode fibers (MMFs) has experienced different periods of rise and fall, mostly driven by their pros and cons in several applications, particularly in optical communications systems \cite{ESSIAMBRE20131}.
On the one side, MMFs are capable of propagating multiple transverse modes, which enables a substantial increase of the information capacity of  fiber-optic links, based on the mode-division-multiplexing technique \cite{richardson2013space}. Furthermore, thanks to their large core area, MMFs deliver higher energies with respect to singlemode fibers, which is advantageous for developing novel high-power fiber laser architectures \cite{wright2017spatiotemporal}.
On the other hand, large modal temporal dispersion has prevented the use of MMFs in long distance high-bit rate transmissions. Moreover, since the multitude of initially excited modes spatially overlaps, the beam quality at the output of MMFs is severely degraded by multimode interference (or modal noise), when compared to singlemode fibers. 
As a result, nowadays singlemode fibers are almost exclusively employed.

In recent years, interest in the use of MMFs has returned within coherent optical communication systems, thanks to their capability for pre or post-compensation of linear modal dispersion effects \cite{richardson2013space,Mecozzi:12,mumtaz2012nonlinear}. Moreover, several novel nonlinear optical effects have been explored \cite{Wright2015R31,Picozzi2015R30,krupa2019multimode,mangini2020spiral}. Remarkably, it has been shown that the Kerr effect can be usefully exploited, in order to provide a bell-shape to the beam profile at the output of graded-index (GRIN) fibers, which permits a significant beam quality and brightness improvement \cite{KrupaPRLGPI}. This effect has been dubbed as spatial beam self-cleaning (BSC)
\cite{krupa2017spatial}. BSC has been demonstrated in several configurations, either in the normal \cite{KrupaPRLGPI,krupa2017spatial,Lopez-Galmiche2016,liu2016kerr,Krupa:16,WrightNP2016,GuenardOpex,GuenardOpex2,9031306, PhysRevA.97.043836,Podivilov2019,Fabert2020} or in the anomalous dispersion regime \cite{leventoux:20,Wu:21,zitelli2021single}, with input pulse duration ranging from nanoseconds to femtoseconds.
Many of these studies have investigated the physical mechanism behind BSC. It is generally accepted that BSC relies on nonlinear coupling among the multiple modes of GRIN MMFs, which leads to a modal redistribution of the input beam energy. Moreover, the robustness of the BSC effect indicates that a sort of equilibrium distribution is established as a result of beam propagation in the MMF. This has led to a description of BSC in a thermodynamics framework. The bell-shaped beam that is spontaneously generated suggests that a prevailing population of the fundamental mode is reached. However, the size of self-cleaned beams was measured to be generally wider than that of the fundamental mode of the MMF \cite{krupa2019multimode}. In addition, the output bell-shaped beam is typically sitting on a wide low-power background. This indicates that higher-order modes (HOMs) also contribute in determining the observed output beam shapes. In this context, it was pointed out that BSC can be seen as a thermalization phenomenon, where the temperature ($T$) is fixed by the coupling condition (CC) of the input laser beam into the fiber \cite{wu2019thermodynamic,Wise2020cleo,pourbeyram2020direct}. 

\begin{figure}[!ht]
\centering\includegraphics[width=7cm]{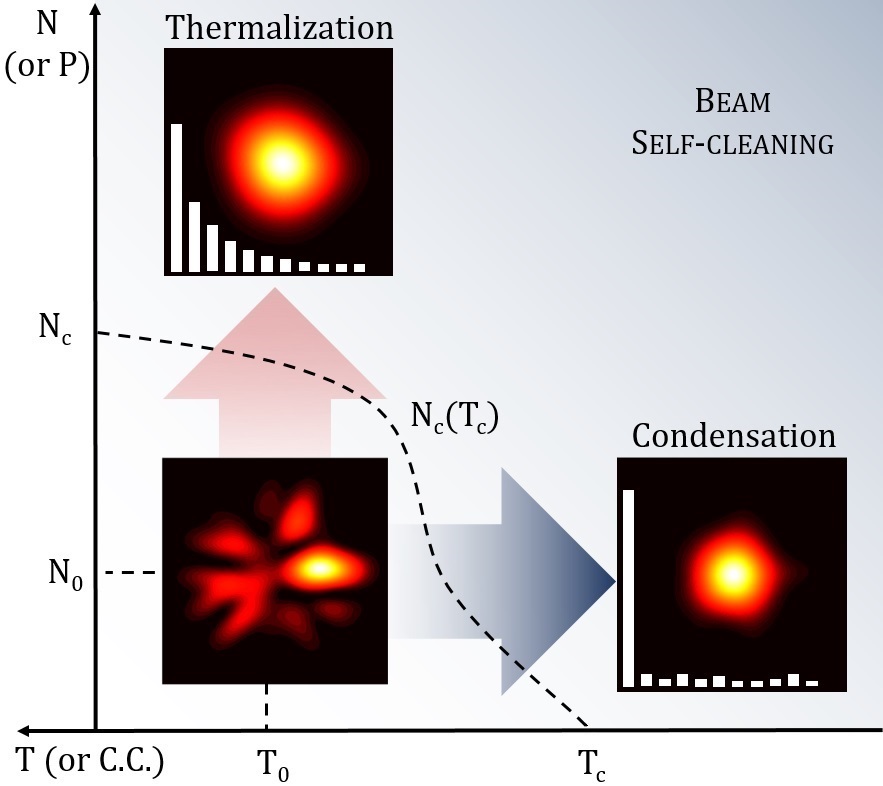}
\caption{Optical phase diagram illustrating orthogonal paths to describe BSC as a thermalization or a condensation process, respectively. The diagram is split in two zones. Below the dashed line (which indicates the condition for observing BSC), optical beams are in a disordered phase (speckled beams). Whereas self-cleaned beams only exist in a certain region (delimited by a dashed black curve) of temperature (or coupling conditions, $CC$) and number of particles (or input power $P$). When passing from one zone to the other by a vertical \emph{phase transition} (red arrow), i.e., by varying $N$ while keeping constant $T=T_0$, BSC can be described as a thermalization of fiber modes. The latter follows the RJ distribution, and HOMs have a non-negligible population (white bars indicate the equilibrium modal occupancy). On the other hand, when keeping fixed $N=N_0$ and decreasing temperature below the critical value (horizontal \emph{phase transition} or grey arrow), BSC can be described as a wave condensation phenomenon, i.e., only the the fundamental mode has a macroscopic population.}
\label{didattica}
\end{figure}

Given that BSC manifests itself when the power of the laser beam that is injected into the fiber overcomes a certain threshold value, one might think of BSC as a sort of 
(quantum) Bose-Einstein condensation of classical waves \cite{aschieri2011condensation}. This is only partially true: as a matter of fact, BSC can be associated to a wave condensation phenomenon, as Baudin and coworkers have recently demonstrated in \cite{baudin2020classical}. However, in that work, it has been shown that at a \emph{fixed} input laser power, BSC can be achieved when the temperature $T$ drops below a certain critical value, i.e., only when the input CC is favorable enough (cfr, the horizontal arrow in Fig.\ref{didattica}). 
On the other hand, in the thermodynamic description the input laser power plays the role of the number of particles ($N$), and it is not related to temperature. Therefore, tuning of the input power does not necessarily trigger wave condensation. As a matter of fact, a thermodynamic system can reach thermal equilibrium even at temperatures which are higher than the condensation temperature. Of course, the input power must be sufficiently high for allowing significant degenerate four-wave-mixing (FWM) among the modes or, in other words, $N$ must be large enough to reach the ergodicity condition \cite{wu2019thermodynamic}.

On the other hand, whenever BSC is observed by increasing the input power (or $N$) for a given input CC (or fixed temperature), the effect can be described as an example of wave thermalization (cfr, the vertical arrow in Fig.\ref{didattica}): here we follow this approach. At thermodynamic equilibrium, the fiber mode distribution obeys the Rayleigh-Jeans (RJ) law, in agreement with the thermodynamic analysis of classical systems comprising a finite number of modes \cite{wu2019thermodynamic}.
In a nutshell, thermalization of the mode distribution is a necessary condition for condensation, but not viceversa. 
As sketched in Fig. \ref{didattica}, the mode distributions corresponding to wave condensation or thermalization are generally different. Because the former involves a dominant (or macroscopic) population of the fundamental mode only, accompanied by energy equipartition among all HOMs. Whereas the latter is characterized by a RJ mode distribution that, although the fundamental mode is always the most populated, may also contain a macroscopic contribution from the adjacent low-order modes. 

Note that, by properly adjusting the input beam coupling conditions 
or by means of wavefront shaping, BSC into the $LP_{11}$ \cite{Deliancourt:19} or even several HOMs \cite{Deliancourt:19b}  could be observed. However, experiments and numerical simulations reported in the supplementary of \cite{Fabert2020} indicate that such HOMs are transient states, which ultimately decay into a mode distribution dominated by the fundamental mode. In these case, the input beam power that is required for thermalization can be higher than the threshold of competing nonlinear effects, such as intermodal FWM and stimulated Raman scattering, which deplete the input beam.
From an experimental point of view, a thermodynamic analysis of BSC requires the use of mode decomposition (MD) techniques for a proper analysis of the output beam mode populations. Several MD methods have been proposed in the literature. Conceptually, they can be divided into different groups, depending on the foundation lying beneath them: genetic algorithms \cite{li2017multimode}, adaptive optics \cite{schulze2012wavefront}, digital holography with spatial light modulators (SLM) \cite{flamm2012mode}, mode-cross correlation analysis \cite{schimpf2011cross}, and beam characterization in terms of spatiotemporal \cite{pariente2016space} or spatial-spectral \cite{nicholson2008spatially} distributions. 

In this work, we present a comprehensive and yet simple description of the phenomenon of BSC within the thermodynamic framework. We derive a mathematical model starting from statistical mechanics, pointing out the dependence of the equilibrium RJ distribution on the radial and azimuthal properties of GRIN fiber modes. In particular, we point out that, if the modes are properly sorted, the RJ distribution can be described as a sequence of sub-equilibrium distributions. It turns out that the modes which have radial symmetry are the most populated in the occurrence of BSC. Whereas, the mode occupancy vanishes when the spatial (azimuthal) complexity grows. 
Theoretical predictions are compared to an extensive MD experimental study, based on the holographic method \cite{gervaziev2020mode}. We investigate BSC when operating with pulses ranging from the femto- to the pico- and nanosecond regime. Our results show that BSC can be reached, largely independently of the input pulse duration, and that the parameters of the ouput RJ distribution can be tuned by acting on the laser-fiber injection condition, or CC. The obtained reconstructions of the spatial profile of the beam at the fiber output turn out to be remarkably similar to the measured patterns, which proves the power and the accuracy of our MD method. Finally, we experimentally verify the statistical mechanics conservation laws, which are at the basis of our theoretical model.
\section{Theory}
\begin{figure}[!ht]
\centering\includegraphics[width=13.3cm]{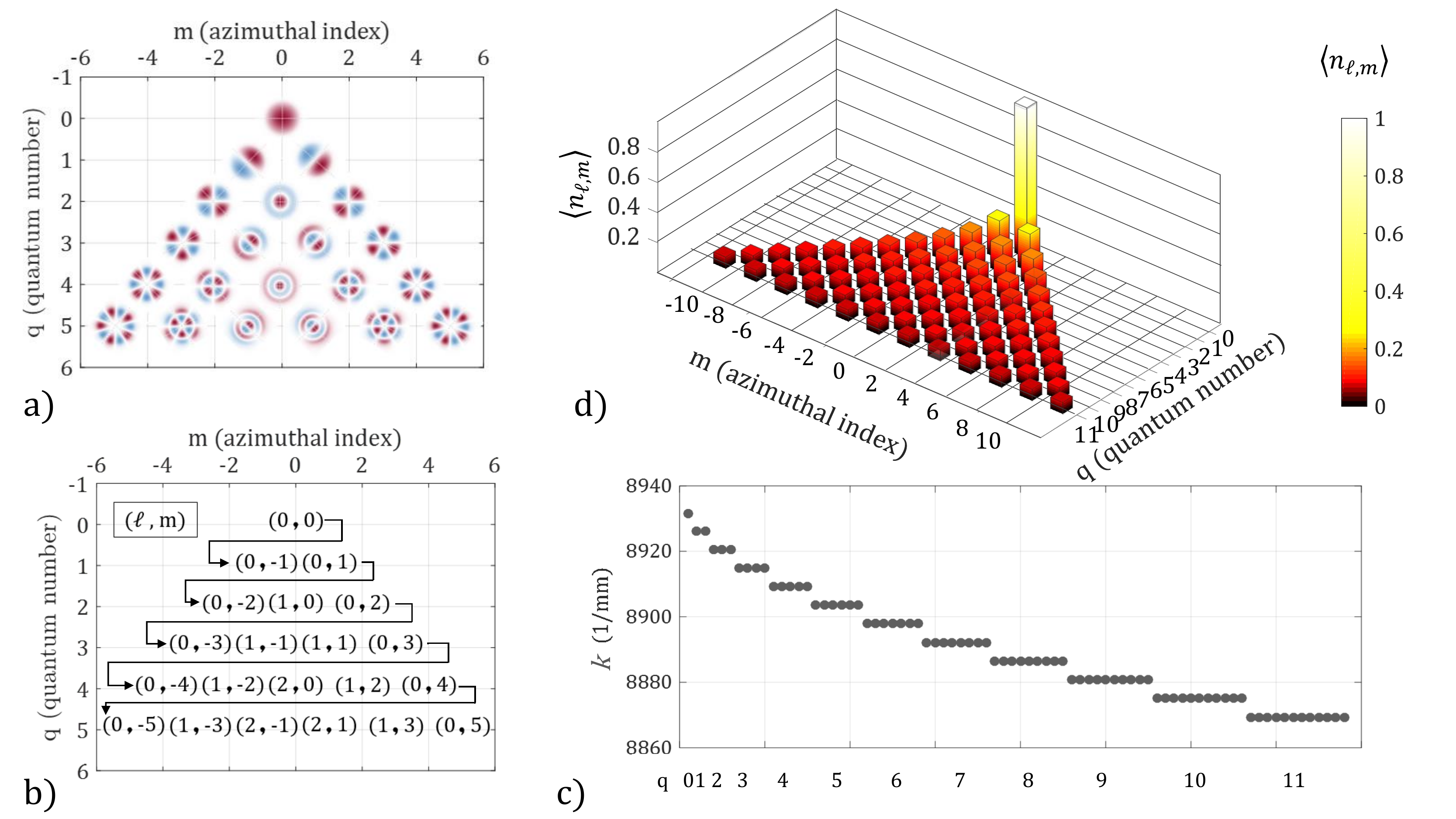}
\caption{a) Illustration of the GRIN MMF modes displayed in the ($q,m$) plane. b) Relative mapping of the indexes ($\ell,m$). c) GRIN fiber modes momentum vs. modes index sorted by quantum number $q = 2\ell+|m|$. d) RJ distribution at the equilibrium in the ($q,m$) plane, when setting $N = 1$, $\mu=-60$ $mm^{-1}$, and $T=0.3$  $mm^{-1}$. The values of $k_{0,0}$ and $k_{SSI}$ are calculated starting from the nominal values of numerical aperture and relative core-cladding index difference provided by the fiber manufacturer.
}
\label{fig-theo}
\end{figure}

Here, we propose a semi-classical approach to determine the mode equilibrium distribution at the occurrence of BSC. In GRIN fibers, modes are conventionally defined in terms of a pair of integer numbers ($\ell,m$), which indicate their radial and azimuthal properties, respectively. As a remarkable example, we may consider Laguerre-Gauss functions as the modal basis, which will be the case for our MD experiments (see Supplementary Materials). In our derivation, we also refer to the quantum number $q = 2\ell + |m|$, so that $m = -q, -q+2, ..., q-2,q$. An illustration of GRIN fiber modes and their relative sorting is provided in Fig.\ref{fig-theo}a. In Fig.\ref{fig-theo}b, we report the couple ($\ell$,m) which is associated to each mode in Fig.\ref{fig-theo}a. The upper limit of $q$, which we dub $Q$, is defined by the fiber cut-off, i.e. by the MMF geometrical parameters and the laser wavelength \cite{mangini2021experimental}. The presence of a frequency cut-off is a fundamental condition for avoiding the divergence of the entropy at low temperatures, which is referred to as \emph{ultraviolet catastrophe} in gas thermodynamics \cite{aschieri2011condensation}. For the sake of simplicity, in Fig.\ref{fig-theo}a we show modes up to $q = 5$. The dynamics of the system of modes can be described by means of an Hamiltonian operator, which only takes into account the linear kinetic term:
\begin{equation}
    H = c \sum_{\ell,m} p_{\ell,m} n_{\ell,m},
    \label{hamilton}
\end{equation}
where $p_{\ell,m}$ and $n_{\ell,m}$ are the momentum and the occupancy of mode ($\ell,m$), and the $c$ speed of light in vacuum. As it has been shown that BSC mainly involves a purely spatial dynamics \cite{krupa2017spatial,PhysRevA.97.043836}, we neglect all temporal effects, and consider monochromatic waves only. 
Still, it is worth to underline that, albeit negligible in a first approximation, the time dimension is needed for BSC, since time-less systems prevent any entropy variation in the presence of the sole FWM \cite{laegsgaard2018spatial}. In other words, FWM by itself is unable to break the coherence of continuous waves (CWs). BSC of CWs can be theoretically achieved if an incoherent spatial process, such as random mode coupling, assists the FWM \cite{fusaro2019dramatic}.
It is worth to noting, anyway, that an experimental demonstration of BSC of CW laser beams has never been reported in the literature, so far.

In the framework of the grand canonical ensemble, we impose that the total energy of the system and the number of particles are conserved, i.e., the Hamiltonian (\ref{hamilton}) is constant and
\begin{equation}
 \sum_{\ell,m} n_{\ell,m} = N.
 \label{number}
\end{equation}
The condition (\ref{number}) holds as long as any dissipative effects can be neglected, which is always the case in practical demonstrations of BSC.
Dubbing $\beta$ and $\mu$ the Lagrange's multipliers corresponding to each conservation law, the partition function then reads \cite{lifshitz2013statistical}: 
\begin{equation}
 Z = \frac{1}{(2\pi\hbar)^{3N}}\cdot\frac{1}{N!} \int e^{-\beta (H-\mu N)} \prod_{\ell ', m'} dn_{\ell ',m'},
\end{equation}
where $\hbar$ is the reduced Planck constant, and the average occupation of mode ($\ell$,m) can be written as:
\begin{equation}
 \langle n_{\ell,m}\rangle = \frac{1}{Z} \cdot \frac{1}{(2\pi\hbar)^{3N}}\cdot\frac{1}{N!} \int n_{\ell,m} e^{-\beta (H-\mu N)} \prod_{\ell ', m'} dn_{\ell ',m'} = \frac{1}{\beta (p_{\ell,m}-\mu N)}.
 \label{RJ-1}
\end{equation}
Eq. (\ref{RJ-1}) is the RJ distribution, which is reached at thermal equilibrium under the hypothesis of the ergodic theorem. The latter holds whenever $N$ is sufficiently high for ensuring a complete exchange of energy among the modes via FWM processes, so that the multimode system explores its whole phase space. Experimentally, this condition reflects the presence of a threshold of the input laser power for achieving BSC. It is important to underline that the ergodicity condition can also be matched, at least in principle, by extending the evolution ''time'' of the system, i.e., by increasing the fiber length. It is well-known, in fact, that the power threshold for BSC decreases as the fiber length increases \cite{krupa2017spatial}. Very often, in fact, people tune the input power as an equivalent cut-back experiment when dealing with nonlinear mode coupling, e.g., when studying soliton self-mode conversion in MMFs \cite{rishoj2019soliton}.
Nonlinear mode coupling is particularly effective in GRIN fibers, which, thanks to their parabolic profile of the core refractive index, forces the beam to periodically shrink after a propagation distance $z_{SSI}$, where SSI stands for spatial-self imaging \cite{hansson2020nonlinear}. This peculiarity originates from the fact that the modes of of GRIN fibers have equispaced momenta, i.e.,
\begin{equation}
    p_{\ell,m} = \hbar k_{\ell,m} = \hbar k_{0,0} - (2\ell + |m|) \hbar k_{SSI},
    \label{momentum}
\end{equation}
where $k_{0,0}$ is the propagation constant of the fundamental mode, which can be calculated starting from the numerical aperture and the core size (usually provided by the fiber manufacturer), and
\begin{equation}
    k_{SSI} = \frac{2\pi}{z_{SSI}}.
\end{equation}
Interestingly, $k_{SSI}$ can be computed by knowing the core and the cladding refractive index; alternatively, it can be derived by directly measuring $z_{SSI}$ through the material defect luminescence \cite{mangini2020multiphoton,mangini2021experimental}. In Fig.\ref{fig-theo}c, we display the modes' momenta, where the modes are sorted by the quantum number $q$ by following the lexicographical order with respect to Fig.\ref{fig-theo}b. Finally, one can calculate the average number of particles in mode ($\ell,m$) as
\begin{equation}
    \langle n_{\ell,m}\rangle = \frac{K_B T}{\hbar k_{\ell,m}-\mu N},
    \label{RJ}
\end{equation}
where we defined $\beta = 1/K_B T$, as it is conventionally done in thermodynamic systems, where $K_B$ is the Boltzmann's constant. Moreover, $\hbar$ is the Planck's constant, 
while $T$ and $\mu$ are parameters which only depend on the input CC, and need to be determined by MD experiments. As $T$ and $\mu$ are only statistical quantities, which do not correspond to a physical temperature o chemical potential of the fiber, in the following we will set $K_B = \hbar = c = 1$. In this way, $T$, $\mu$ and $H$ have the dimension of the inverse of a length. In Fig.\ref{fig-theo}d, we plot the thermal distribution (\ref{RJ}) in the ($q,m$) space.

\section{Experimental setup}
\begin{figure}[ht!]
\centering\includegraphics[width=13cm]{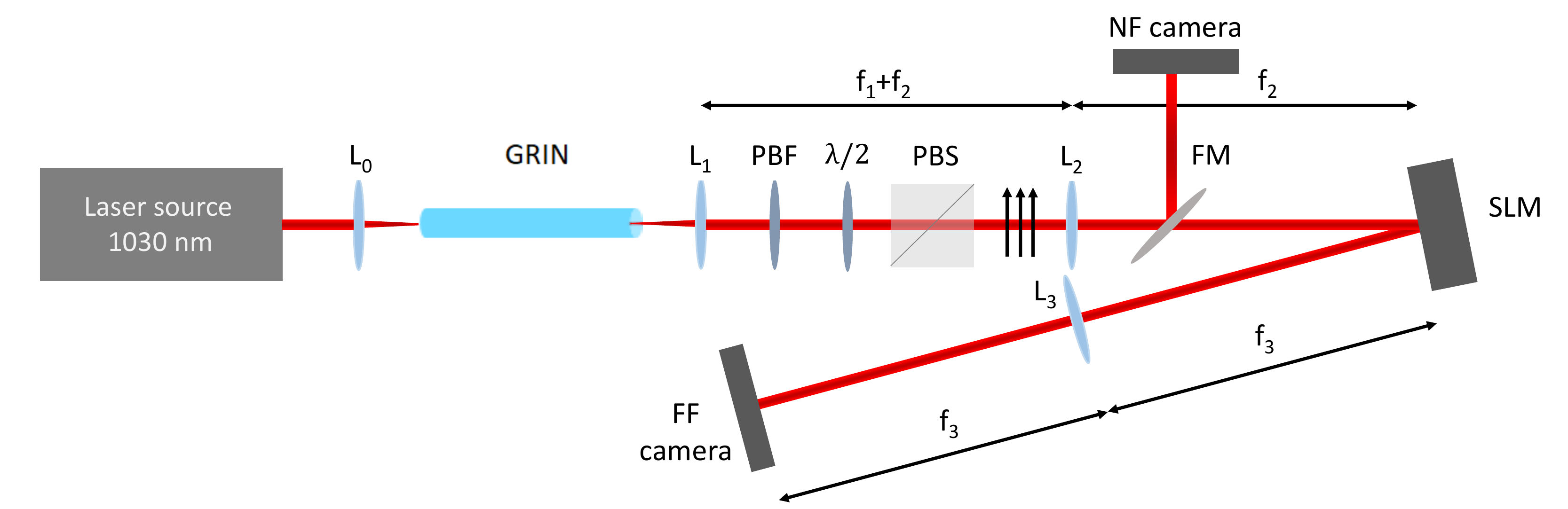}
\caption{Sketch of the experimental set-up. The lenses focal distances are $f_1 = 2.75$ mm, and $f_2 = f_3 = 400$ mm. The arrows indicate the beam polarization direction (horizontal).}
\label{set-up}
\end{figure}

The MD experimental setup that we used to study BSC in GRIN MMFs is shown in Fig.\ref{set-up}. It consists of an ultra-short pulse laser system pumped by a femtosecond Yb-based laser (Lightconversion PHAROS-SP-HP), generating pulses with adjustable duration (by means of a dispersive pulse stretcher), at 100 kHz repetition rate and $\lambda=1030$ nm, and with Gaussian beam shape ($M^2$=1.3). The pulse shape was measured by using an autocorrelator (APE PulseCheck type 2), resulting in a sech temporal shape with pulse width ranging from  174 fs up to 10 ps. As shown in Fig.\ref{set-up}, the laser beam was injected by a positive lens ($L_0$) into the core of the GRIN fiber. The input diameter at $1/e^2$ of peak intensity was measured to be $30$ $\mu$m.
We employed 3 m long standard 50/125 GRIN fibers (GIF50E from Thorlabs), whose core radius, core refractive index along the axis, relative core-cladding index difference, numerical aperture and fundamental mode radius at $\lambda$ = 1030 nm are $r_c=25$ $\mu$m, $n_0=1.472$, $\Delta=0.0103$, $NA = 0.2$ and $r_{0,0} = 6.33$  $\mu$m, respectively. 
The near-field profile at the fiber output is imaged onto an SLM (Hamamatsu LCOS- X15213) by means of two confocal lenses ($L_1$, with $f_1=2.75$ mm and $L_2$, with $f_2=400$ mm). Between those, we placed a bandpass filter (BPF, $1030 \pm 5$ nm), a half-wave plate ($\lambda/2$), and a polarizer (PBS). Our measurement system allows us to avoid the parasitic influence of nonlinear frequency conversions, e.g., provided by Raman scattering or geometric parametric instability \cite{KrupaPRLGPI}, which is detrimental for our MD reconstruction algorithm. As a matter of fact, the phase pattern on the SLM for the profile reconstruction algorithm must be chosen at a given wavelength. We could also tune the intensity of the beam reaching the SLM by means of the $\lambda/2$.
A flip mirror (FM) is used for imaging the near field profile at the fiber output facet onto an IR camera (NF, Gentec Beamage-4M-IR). Images acquired in this way were used as a reference, in order to check the quality of the reconstruction made by the MD algorithm.
At last, a convex lens ($L_3 = 400$ mm) projects the field reflected by the SLM onto a second camera (FF camera). The lens is placed in the middle between the SLM and the camera, so that both these objects are at its focal distance.
Finally, the beam average power both at the input and the output of the fiber was measured by a photo-diode power meter (Thorlabs). In order to explore the nanosecond pulse regime, we carried out analogous MD experiments by using a Nd:YAG laser emitting pulses of 0.435 ns duration and 1064 nm of wavelength, at 1 kHz of repetition rate.

\section{Results}
Since the values of the intensity profiles which are detected by the cameras are given in arbitrary units, our setup only allows for evaluating the relative values of mode occupancy, i.e., the power fraction values, normalized by the total beam power. Therefore, in the following of our analysis we set $N=1$, so that $n_{\ell,m}$ represents the occupancy fraction of mode ($\ell,m$). Moreover, we limited our experimental analysis to the first 78 modes with the highest values of momentum, i.e., we only considered modes with $q\leq11$ (see Fig.\ref{fig-theo}c). We found out that such number was a good compromise between the time convergence of the MD algorithm, and the quality of the near-field reconstructions. Further details about the MD method are presented in the Supplementary section.

In Fig.\ref{main}, we report results obtained when employing 7.6 ps input laser pulses. Specifically, what we show in Fig.\ref{main}a is a 3D histogram of the experimental mode power fractions, for different values of the input beam peak power ($P_p$). Whereas, in Fig.\ref{main}b, we illustrate the experimental evolution, as a function of $P_p$, of the sum of the power fraction of all modes with the same quantum number. This figure shows that the occupancy of the fundamental mode (which is the only mode with $q=0$) grows larger when $P_p$ is increased, indicating the progressive occurrence of BSC, accompanied by thermalization of the mode distribution. In particular, it can be seen that at the lowest input power value of 110 W, the CC is such that the sum of the occupancy of modes with $q=1$ is higher than that of the fundamental mode. The latter progressively becomes dominant, when increasing the input power, up to the occurrence of BSC.
In Fig.\ref{main}c, we compare the experimental power fraction values (blue bars) with the theoretical RJ distributions (red dashed lines). As it can be seen, as $P_p$ increases, the experimental distribution progressively approaches the theoretical RJ distribution. This can be quantitatively appreciated in Fig.\ref{main}d, where we plot the root-mean-square error (RMSE) of the observed mode distributions with respect to the RJ distribution, as a function of $P_p$. Finally, insets of Fig.\ref{main}c demonstrate the excellent quality of our MD method. Here, we show that the measured output near field intensities (images in the left column) are impressively similar to their MD reconstructions (images in the right column), for all input power values. 

\begin{figure}[ht!]
\centering\includegraphics[width=13.3cm]{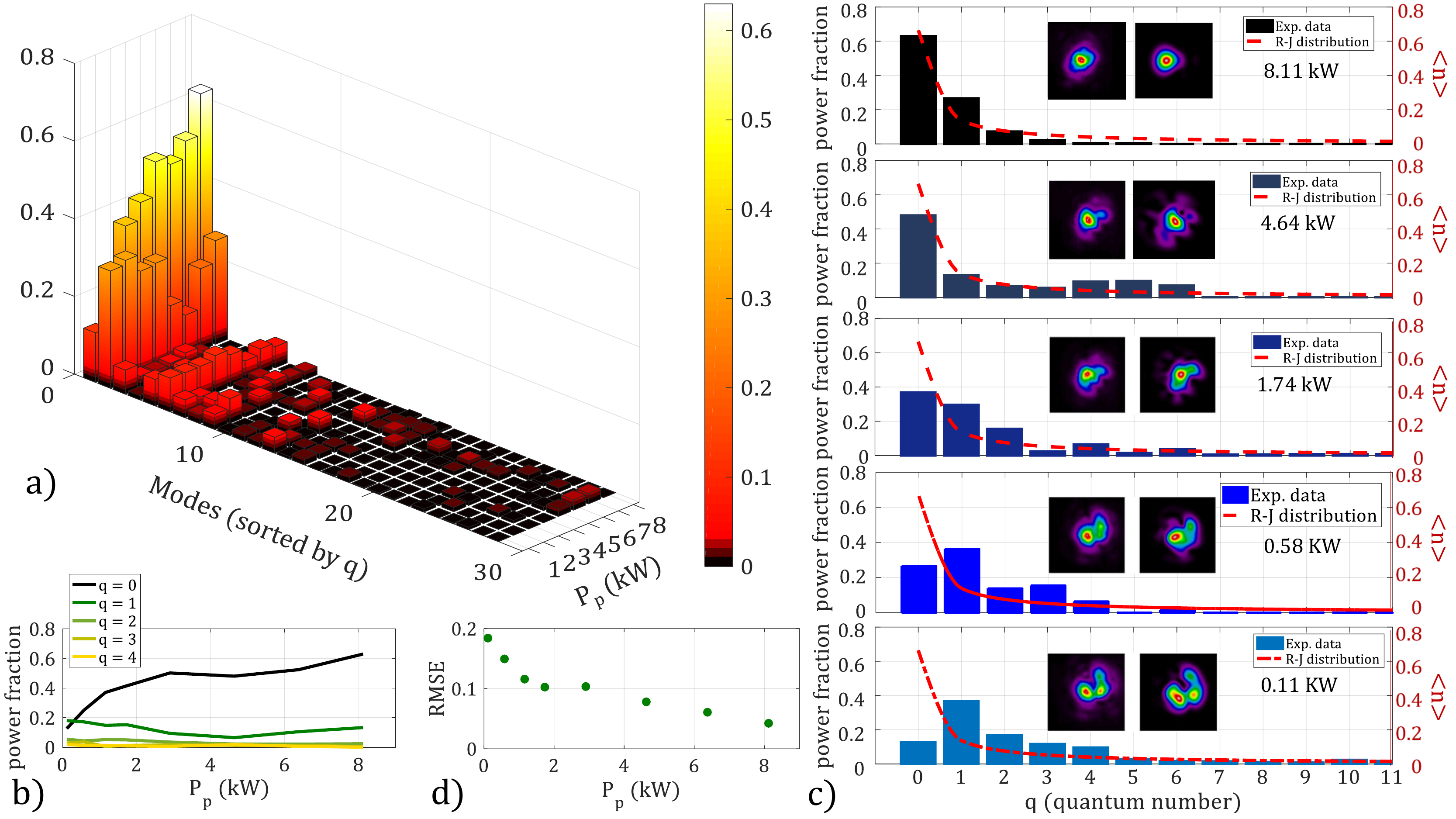}
\caption{MD analysis for 7.6 ps of input pulse duration. a) Mode distribution (with the same sorting as in Fig.\ref{fig-theo}c) for different values of the input peak power. b) Experimental value of the power fraction of the fiber modes grouped by $q$, vs. input power $P_p$. c) Details of the reconstruction in 2D plots. The images in the inset of each graph represent the measured (left) and reconstructed (right) output beam profile. The solid line is the RJ distribution that fits the experimental values at the highest value of input power (8.11 kW). The fitting parameters are $\mu = -63.97$ $mm^{-1}$ and $T = 0.92$ $mm^{-1}$. d) Root-mean-square error (RMSE) of the fitting curve in b), when varying $P_p$.\\
}
\label{main}
\end{figure}

\subsection{BSC stability vs input pulse duration}
\begin{figure}[ht!]
\centering\includegraphics[width=13.3cm]{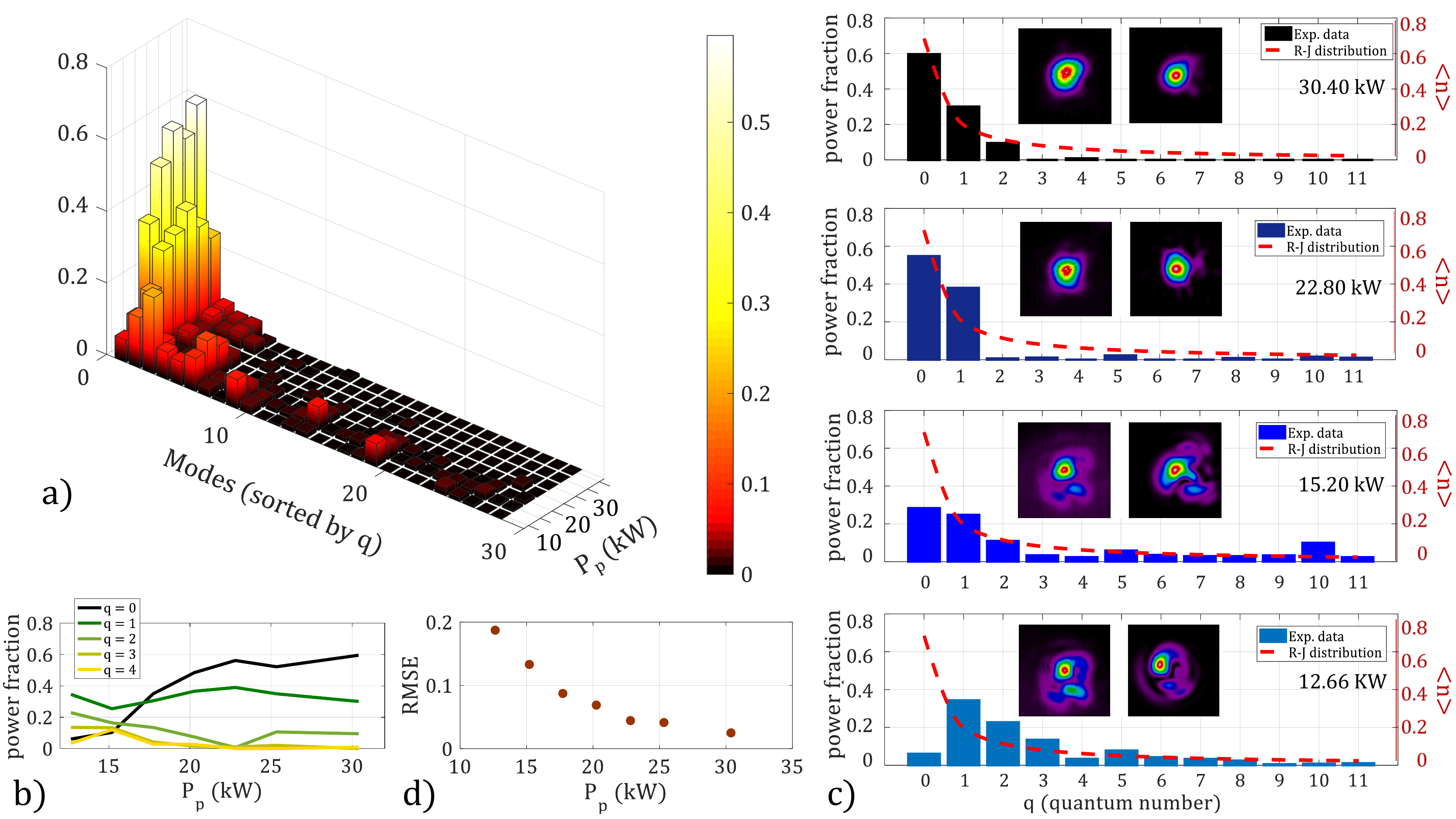}
\caption{Same as Fig.\ref{main} for pulses of 174 fs of duration. The CC is the same as in Fig.\ref{main}. The fitting parameters are $\mu = -63.96$ $mm^{-1}$ and $T = 1.01$ $mm^{-1}$.}
\label{main-fs}
\end{figure}

We carried out MD experiments with pulses whose durations ranging from hundreds of femtoseconds up to 0.5 nanoseconds. In all cases, we kept the same fiber length, whereas we used the Nd:Yag laser that emits pulses of 0.5 ns of duration. 
In Fig.\ref{main-fs}, we present the MD analysis based on 174 fs input pulses.
This figure allows for further appreciating the thermalization nature of BSC. We could observe, in fact, that the fundamental mode power fraction has a non-monotonic behavior when increasing $P_p$, although the RMSE with respect to the RJ distribution keeps decreasing. At $P_p \simeq 26$ kW, the power fraction $\langle n\rangle$ of the fundamental mode (with $q=0$) has a local minimum (see Fig.\ref{main-fs}a and b).
It is important to underline that MD experiments of Fig.\ref{main} and Fig.\ref{main-fs} with different pulse duration were carried out with the same laser-fiber CC. Our laser allows, in fact, to vary the pulse duration, while the pulse energy is kept constant. Therefore, we could maintain the experimental set-up untouched, thus avoiding any modification of the injection conditions.

\begin{figure}[ht!]
\centering\includegraphics[width=13.4cm]{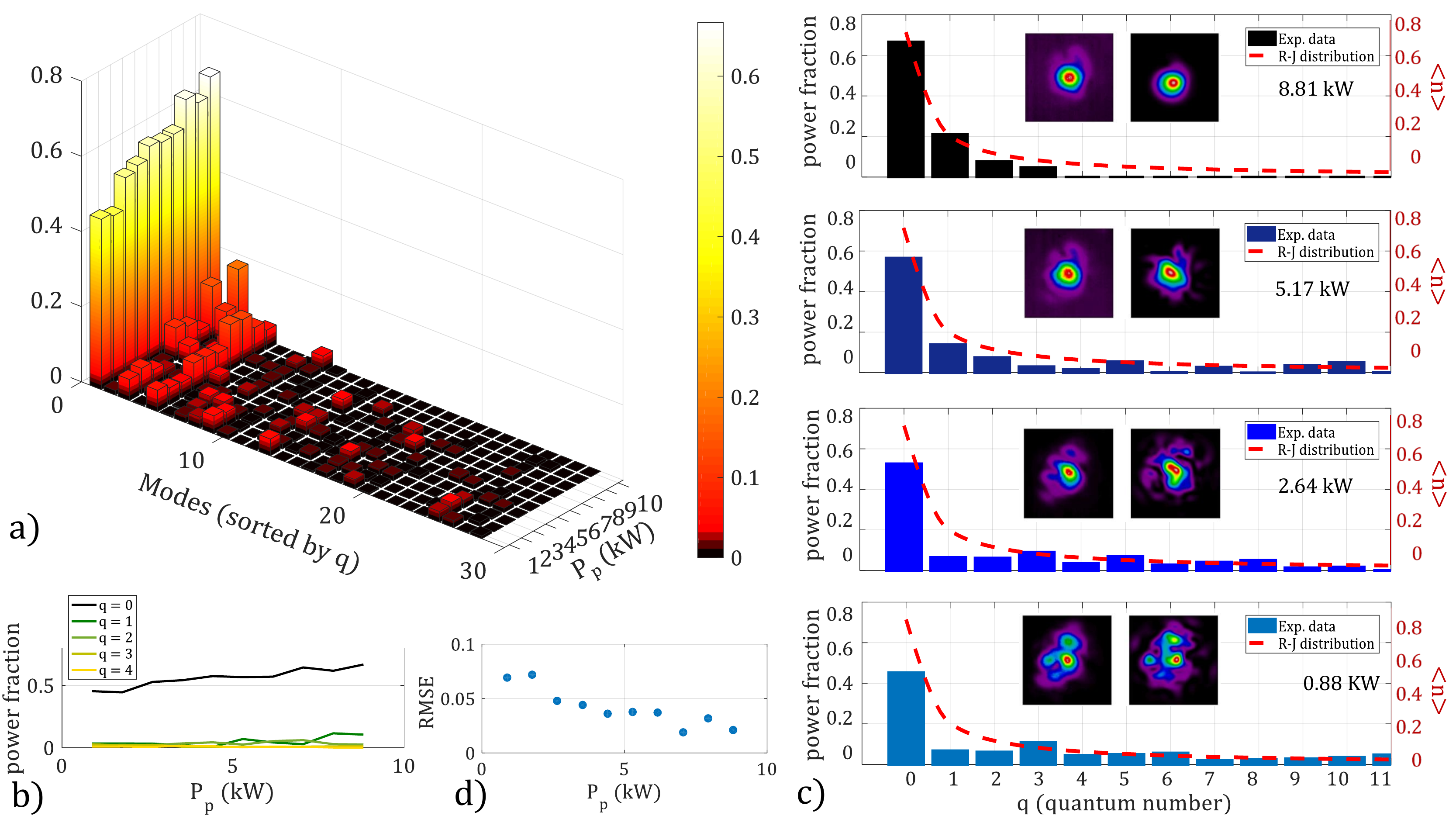}
\caption{Same as Fig.\ref{main} for pulses of 1 ps of duration, and different CC. The fit parameters are $\mu = -64.26$ $mm^{-1}$ and $T = 1.08$ $mm^{-1}$.}
\label{main-ps}
\end{figure}

\begin{figure}[ht!]
\centering\includegraphics[width=13.3cm]{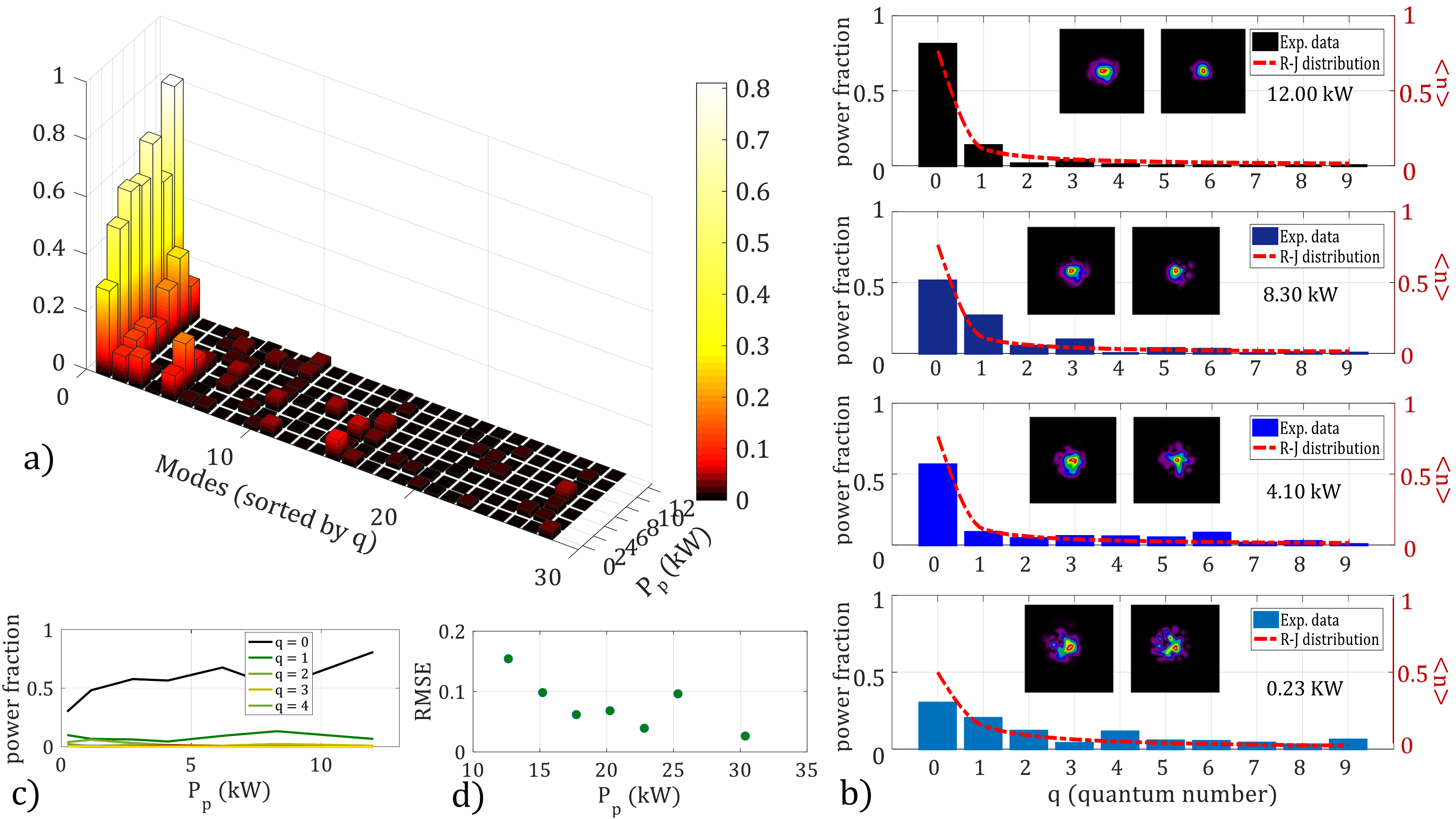}
\caption{Same as Fig.\ref{main} for pulses of 435 ps of duration and different injection conditions. The fit parameters are $\mu = -51.86$ $mm^{-1}$ and $T = 0.72$ $mm^{-1}$.}
\label{main-ns}
\end{figure}

We found almost identical values of the parameters $\mu \simeq 64$ $mm^{-1}$ and $T\simeq 1$ $mm^{-1}$ in these two cases. In agreement with theoretical predictions, in fact, $T$ and $\mu$ do not depend on the pulse duration, and they can be varied only by changing the CC \cite{baudin2020classical}. This is the case of the MD analysis which is reported in Fig.\ref{main-ps} and \ref{main-ns}, where we found different values of $T$ and $\mu$ with respect to those in Fig.\ref{main} and \ref{main-fs}.
Besides their different laser-fiber CC, the results reported in Fig.\ref{main-ps} and \ref{main-ns} have been obtained awith different pulse duration of 1 picosecond, thus experimentally confirming that BSC can be achieved with pulses ranging from the femtosecond up to the sub-nanosecond regime. 

\subsection{Radial modes distribution}
\begin{figure}[ht!]
\centering\includegraphics[width=12cm]{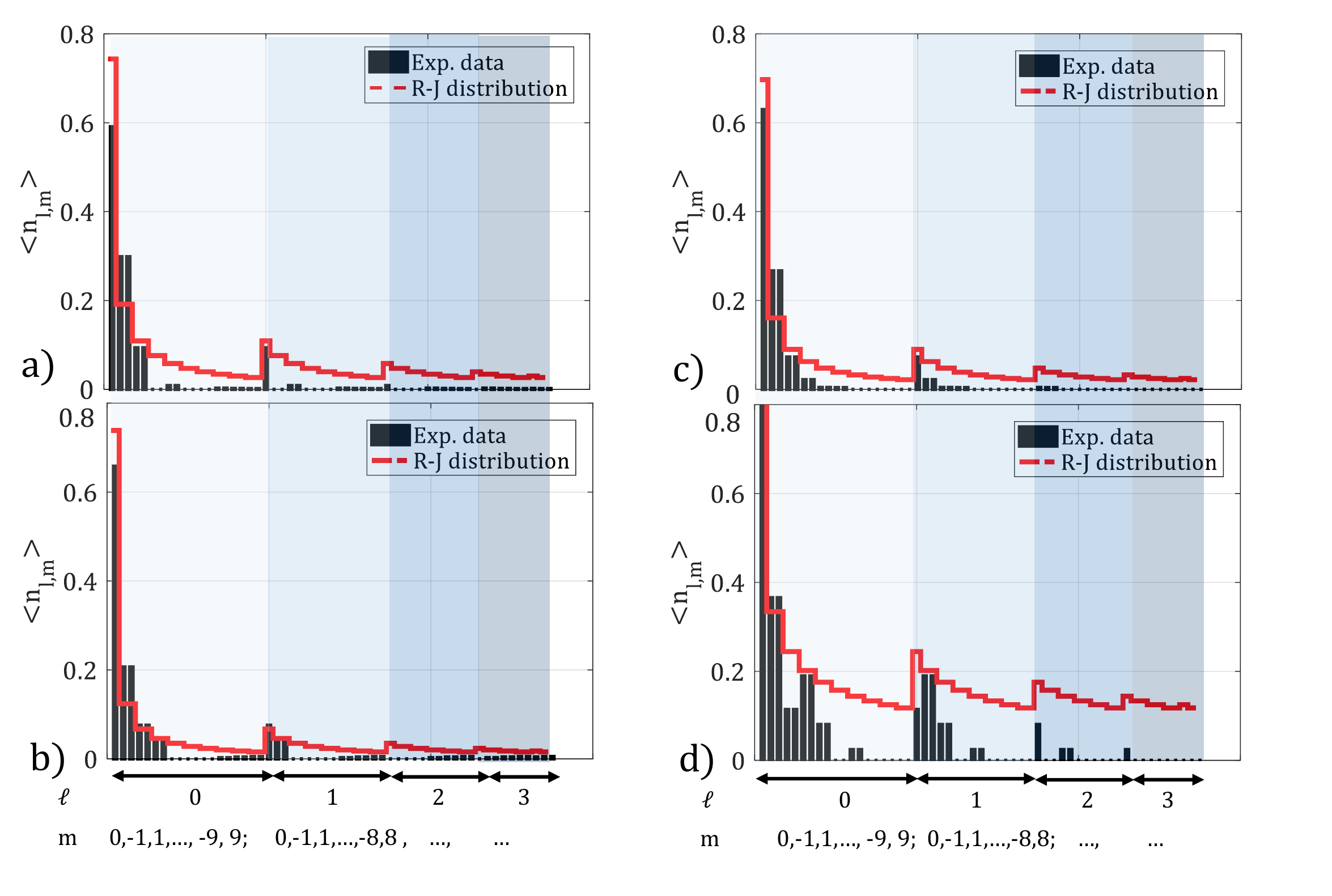}
\caption{Illustration of the mode decomposition results when sorting the modes by their radial index $\ell$ for input pulse duration of 174 fs (a), 1 ps (b), 7.6 ps (c), and 435 ps (d). Black bars represent the experimental values of the power fraction of each mode, while the red dashed line is the theoretical RJ distribution.}
\label{radial}
\end{figure}

So far, when showing modal distributions, we sorted the modes by their quantum number. This is the most convenient choice, when one limits the analysis to beams which do not carry angular momentum, i.e., when injecting a radially symmetric laser beam at the center of the fiber core, as it occurs in our case. However, our theoretical derivation aims to be more general, since it has been developed on the basis of Laguerre-Gauss modes, which depend on two indexes (radial and azimuthal). Then one may naturally wonder how the equilibrium distribution looks like when sorting the modes by $\ell$ and $m$. This is shown in Fig.\ref{radial}, for all considered values of the input pulse duration.

Fig.\ref{radial} permits to clearly appreciate that the RJ distribution is indeed a sequence of sub-distributions. As a matter of fact, for each value of the radial index $\ell$, the mode having $m=0$ is the most populated one at the equilibrium. Whereas the occupancy $\langle n_{\ell,m\neq0}\rangle$ rapidly vanishes as $m$ grows. We underline that the steps of the red solid curve (the theoretical RJ distribution) when passing from $\ell$ to $\ell+1$ are due to the finite number of modes of the system, which in our case is due to the mode truncation that we made when running our algorithm. However, similar curves, albeit much longer, would have been obtained, had we considered the presence of all modes below cut-off.

When modes are sorted by the radial index, their distribution emphasizes the symmetry with respect to the azimuthal index $m$ of the equilibrium distribution. From a mathematical point of view, this is provided by the presence of the absolute value in the definition of the momentum of the GRIN fiber modes, see Eq.(\ref{momentum}). Thus, according to Eq.(\ref{RJ}), $\langle n_{\ell,m}\rangle = \langle n_{\ell, -m}\rangle$. Experimentally, this is demonstrated by Fig.\ref{radial}. As it can be seen, each couple of bars corresponding to modes with the same value of $\ell$, and opposite signs of $m$ have the same height, again independently of the pulse duration. 


\subsection{Conservation laws verification}
Further investigating the symmetry of mode occupancies with respect to $m$ highlights an important feature of BSC: the conservation of the mode parity ($M$), which is defined as:
\begin{equation}
    M = \hbar \sum_{\ell,m} m \cdot n_{\ell,m}.
    \label{parity}
\end{equation}
The conservation of $M$ is related to the conservation of the total angular momentum, which is intrinsically provided by the formulation of the Hamiltonian (\ref{hamilton}). In our derivation, in fact, we did not consider any contribution to the energy that may arise from the angular momentum carried by each mode, i.e., analogously to rotational kinetic energy of classical mechanics. Besides, we studied the power dependence of the Hamiltonian and the total number of particles, whose conservation laws are at the basis of our theoretical model.

\begin{figure}[ht!]
\centering\includegraphics[width=16.5cm]{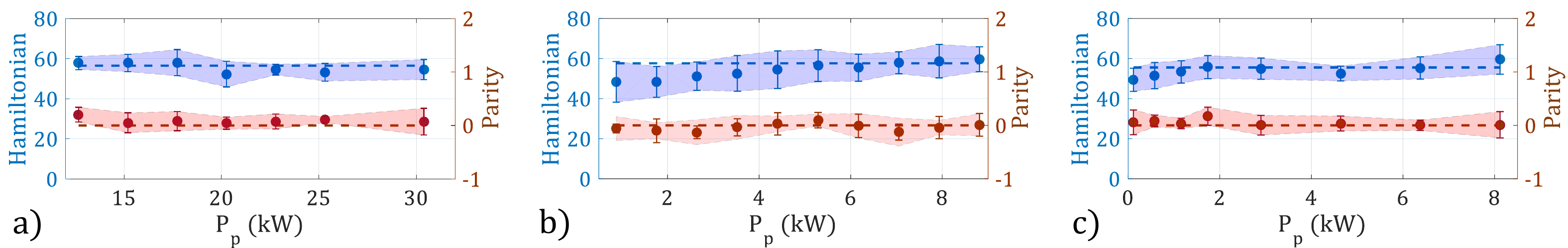}
\caption{Conservation of the Hamiltonian $H$ and of parity $M$, for an input pulse duration of 174 fs (a), 1 ps (b) and 7.6 ps (c), respectively. The error bars are estimated by considering all of the reconstructions of the output beam near-field at each input power (see the MD method section in the Supplementary Materials). The horizontal dashed lines are obtained from the RJ distribution that fits the experimental data in the BSC regime.}
\label{conservation}
\end{figure}

In Fig.\ref{conservation}, we show that both the Hamiltonian and the parity remain nearly constant when varying the input peak power, for all input pulse durations. Specifically, we found that $H_{exp}\simeq 55.3$ $mm^{-1}$ and $M_{exp} \simeq 0.04$, $H_{exp}\simeq 55.4$ $mm^{-1}$ and $M_{exp} \simeq -0.02$, and $H_{exp}\simeq 54.1$ $mm^{-1}$ and $M_{exp} \simeq 0.01$ $mm^{-1}$, at 174 fs, 1 ps, and 7.6 ps of pulse duration, respectively. These results can be compared to the values of $H$ and $M$ that are calculated by substituting the RJ distribution which fits the experimental data in Eq.(\ref{hamilton}) and Eq.(\ref{parity}). We found that $H_{RJ} =  56.7$ $mm^{-1}$, $H_{RJ}\simeq 57.6$ $mm^{-1}$, $H_{RJ}\simeq 56.1$ $mm^{-1}$ at 174 fs, 1 ps, and 7.6 ps of pulse duration, respectively. Whereas $M_{RJ} = 0$, owing to the above-mentioned symmetry of the RJ distribution.

Finally, the number of particle is well-known to be conserved, since linear absorption is less than 1 dB/km at the wavelength of operation \cite{agrawal2000nonlinear}. The number of particles conservation law would be broken if we were working in different regimes, e.g., when dealing with non-negligible linear or nonlinear absorption. Nevertheless, this would require to reach, on one side, ultraviolet or mid-infrared wavelengths (i.e. for reaching the silica bandgap or triggering phonon resonances) or, on the other, pulse peak powers that are way above the BSC threshold \cite{ferraro2021femtosecond}.

\section{Conclusions}
In this work, we provided a comprehensive description, based on statistical mechanics, of the physical mechanisms behind BSC. Our approach permits a simple derivation, from fundamental principles, of the mode distribution at the thermodynamic equilibrium.
We have compared the theoretical predictions of the BSC mode distributions in GRIN MMFs, with a full experimental characterization, based on a MD holographic technique.
Theory and experiments were found to be in a very good quantitative agreement. Specifically, the mode distribution associated with BSC, that is experimentally observed at the output of a GRIN fiber when increasing the input power, can be well described by a RJ distribution, indicating that optical thermalization is reached. The parameters that define the RJ distribution, i.e., the temperature and the chemical potential, only depend on the spatial coupling conditions of the input beam. Whereas the equilibrium distribution parameters turn out to be largely independent of the input pulse duration. In the thermodynamic analogy, larger input powers mean greater numbers of particles, and a faster approach to an ergodic mixing of all states, or thermalization. 
The thermodynamic equilibrium is reached by ensuring the conservation of the total energy and the number of particles of the multimode system. Furthermore, we experimentally show that the symmetry with respect to the index $m$, which is related to the total angular momentum, is also a conserved quantity.
Finally, our observations reveal that, as a result of thermalization, modes with radial symmetry exhibit a macroscopic occupation, at the expenses of modes with more complex azimuthal characteristics.
The results of our studies provide a significant contribution to the fundamental understanding of an intriguing physical process, and will be of general interest for applications of multimode nonlinear fibers in different emerging technologies, ranging from nonlinear imaging to beam delivery and high-power lasers.

\section*{Acknowledgments}
We acknowledge support from the European Research Council (ERC) under the European Union’s Horizon 2020 research and innovation program (grant No. 740355 and grant No. 874596), the Italian Ministry of University and Research (R18SPB8227), and the Russian Ministry of Science and Education Grant No. 14.Y26.31.0017. M.G. was also supported by RFBR (Grant No. 20-32-90132), S.B. and D.Kh were supported by RSF (Grant No. 21-42-00019).

\section*{Disclosures}
The authors declare that they have no competing financial interests.


\bibliographystyle{unsrtnat}
\bibliography{references}  
\onecolumn
\noindent {\Huge \textbf{Supplementary Materials}}

\section*{Mode-decomposition method}
Here, we summarize the main steps of the MD method, based on digital computer holography \cite{flamm2012mode}. The transverse ($x,y$) beam profile at the output of a GRIN MMF may be represented as the superposition of Laguerre-Gaussian (LG) modes 
\begin{equation}
   U(x,y) = \sum_{\ell,m = 0}^{\infty} \xi_{\ell, m} \cdot \psi_{\ell, m}(x,y),
    \label{eqModesConsist}
\end{equation}
where $\xi_{\ell,m}$ are the normalized complex amplitudes of each mode and $\psi_{\ell,m}$ are LG functions. The main objective of MD is that of searching for the values of $\xi_{\ell,m}$, which are complex numbers that are defined as the scalar product between $U$ and $\psi_{\ell,m}$, i.e.,
\begin{equation}
    \xi_{\ell,m} = \langle\psi_{\ell,m}|U\rangle = \iint_{-\infty}^{+\infty} \psi_{\ell,m}^*(x,y) U(x,y)dxdy.
    \label{eq-xi}
\end{equation}
Actually, as we target the mode occupancy $n_{\ell,m}$, we do not need to know amplitude and phase of $\xi_{\ell,m}$, but only its square modulus, as $|\xi_{\ell,m}|^2 = n_{\ell,m}$.

In our experiments, we measure the far-field (FF) intensity $I$ of the superposition of the phase mask provided by the SLM and $U(x,y)$, opportunely magnified by the lenses $L_1$ and $L_2$ (see Fig.\ref{set-up}). We dub $T$ the transmission function which transforms $U$ as a consequence of $L_1-L_2$ magnification and the SLM phase encoding. Then the field profile on the FF camera is obtained by calculating the Fourier transform (given by the lens $L_3$) to the product of $T$ and $U$, so that $I$ reads as
\begin{equation}
    I(k_x,k_y) = \left| \iint_{-\infty}^\infty T(x,y)U(x,y)\mathrm{e}^{\mathrm{i}(k_xx + k_yy)}dxdy \right|^2,
\end{equation}
where the square modulus comes from the FF camera measurement. In order to distinguish the contribution to $I$ coming from each mode ($\ell,m$), we properly choose several phase patterns to be encoded on the laser beam. For the sake of simplicity, here, we label each phase pattern with the same indexes ($\ell,m$) of the modes. Therefore, we substitute $T \xrightarrow{} T_{\ell,m}$ and $I \xrightarrow[]{} I_{\ell,m}$, so that
\begin{equation}
    I_{\ell,m}(k_x,k_y) = \left| \iint_{-\infty}^\infty T_{\ell,m}(x,y)U(x,y)\mathrm{e}^{\mathrm{i}(k_xx + k_yy)}dxdy \right|^2.
    \label{eq-I}
\end{equation}
With this substitution it looks clear that if the transmission function is chosen such that $T_{\ell,m} (x,y) = \psi_{\ell,m}^* (x,y)$, Eq.(\ref{eq-I}) boils down to the square modulus of Eq.(\ref{eq-xi}) calculated in $k_x = k_y = 0$. Therefore, by properly choosing a set of phase pattern on the SLM, a corresponding set of images captured by the FF camera can be used for calculating all of the $n_{\ell,m}$. For a full explanation of the method, and details about the phase patterns, one can refer to \cite{gervaziev2020mode}.

\subsection*{Estimation of the decomposition error}
In Fig.\ref{conservation}, we show that, when varying the input power, the total angular momentum and the Hamiltonian are constant within the limits of the experimental error. In our report, we fully address the latter to the numerical reconstruction of the mode composition. As a matter of fact, the experimental MD analysis consists of two parts. In the first part, an algorithm simultaneously control the SLM and the FF camera, storing then a set of images, each of which corresponding to a given phase pattern of the SLM. The second part starts from these images for reconstructing the near field (NF) of the beam the output facet of the fiber, by determining the parameters $n_{\ell, m}$, as discussed above. In this second part, the choice of the center ($k_x =0, k_y = 0$) is as pivotal as non-trivial. The images, in fact, are the superposition of the output beam with a series of phase patterns. The identification of a center is not necessarily straightforward, as the $I(k_x,k_y)$ does not own particular symmetries (even at the occurrence of BSC).

We verified that even a single pixel offset in the choice of the center point may lead to the lost of faithful reconstructions. Therefore, we estimate the error bars in Fig.\ref{conservation} by running several time the reconstruction algorithm for a given set of acquired images. We first choose the most faithful reconstruction among the several we made, thus defining the mean value of the experimental Hamiltonian and total angular momentum. Then, we define their error as the difference from the reconstructions obtained when choosing a center point that has a single pixel offset with respect to the chosen one.

\section*{Temporal and spectral effects}
\begin{figure}[!ht]
\centering\includegraphics[width=7cm]{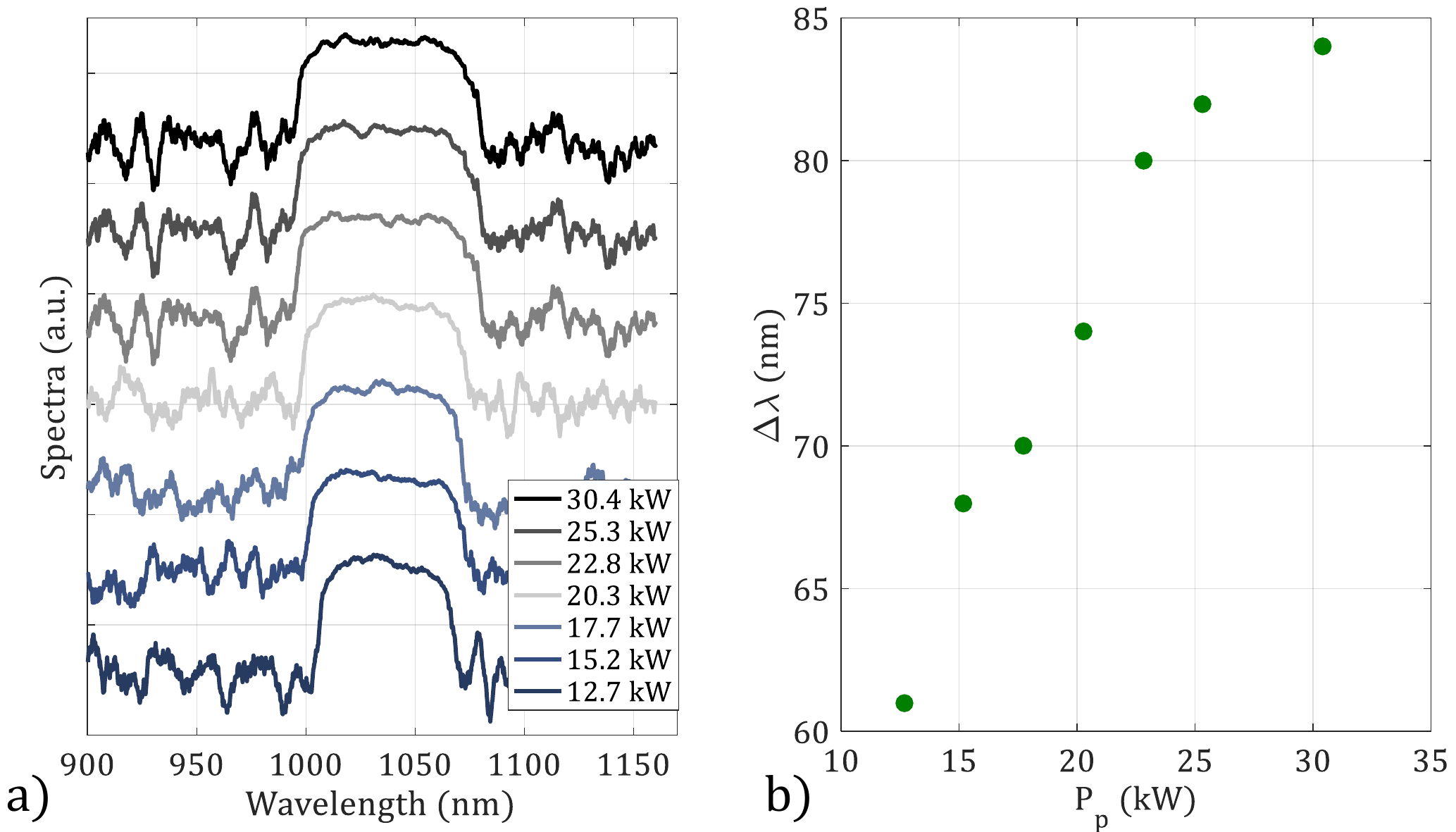}
\caption{a) Output spectrum \textbf{in logarithmic scale} at input peak powers up to 30 kW (BSC power threshold). The input pulse duration is 174 fs. b) Output bandwidth vs. input peak power.}
\label{spectra174fs}
\end{figure}

In the main text we only report spatial properties of the beam at the output of GRIN MMFs. This is because BSC is a spatial phenomenon, which mainly involves spatial dynamics. As we pointed out in the main text, BSC theoretical description can be derived for CWs. Nevertheless, when dealing with pulsed laser sources, one naturally wonders what are the temporal and spectral effects during the propagation inside a nonlinear medium. Here, we give a brief discussion about the role of group velocity dispersion and the spectral broadening. Specifically, we consider the case of the shortest pulses used in our experiments, i.e. that having 174 fs of duration, as temporal and spectral effects are the most pronounced.
When employing intense ultra-short pulses, self-phase modulation produces a spectral broadening of the pulse. Therefore, at the fiber output the spectrum results quite broader than that at the input. In Fig.\ref{spectra174fs}.a, we report the output spectrum at several values of $P_p$. Specifically, the latter ranges from 12 kW up to 30 kW, which is the threshold value for achieving BSC. For a clearer comparison, $P_p$ values reported here are the same of Fig.\ref{main-fs}. Spectra are shown in a log scale in order to emphasize their broadening. 
In Fig.\ref{spectra174fs}b, we show the the dependence on the input peak power of the output pulse full-wave-half-maximum bandwidth ($\Delta\lambda$), calculated from Fig.\ref{spectra174fs}a. As it can be seen, $\Delta\lambda$ reaches the value of 85 nm at the occurrence of BSC. This explains the need of the band-pass filter in the experimental set-up in Fig.\ref{set-up} when working with short pulses. Indeed, if the filter was missing, the near-field camera would detect an incoherent superposition of waves with different wavelength, whose average would be seen as a Gaussian "cleaned" beam. That camera, in fact, detect light with frequencies ranging from the visible to the near-infrared. Furthermore, the SLM can encode a proper phase mask only to a monochromatic source. Thus, in the absence of the band-pass filter, MD cannot be properly done.\\
As a final note, it is worth to underline that the conservation of the number of particles must be verified by measuring the output power before the beam passed through the band-pass filter. For this reason we used a large-band thermal power meter.

\end{document}